\documentclass[11pt]{article}

\usepackage{graphicx}
\usepackage{amsmath}

\usepackage{color}

\title{Increased methane emissions from deep osmotic and buoyant convection beneath submarine seeps as climate warms }

\author{Silvana S. S. Cardoso \\
Department of Chemical Engineering and Biotechnology, \\ University of Cambridge,  
Cambridge CB2 3RA, UK, \\ sssc1@cam.ac.uk \\
Julyan H. E. Cartwright \\
Instituto Andaluz de Ciencias de la Tierra, \\ CSIC--Universidad de Granada, \\ E-18100 Armilla, Granada, Spain and \\
Instituto Carlos I de F\'{\i}sica Te\'orica y Computacional, \\ Universidad de Granada, E-18071 Granada, Spain, \\ julyan.cartwright@csic.es
}

\begin{document}

\maketitle

\emph{
High speeds have been measured at seep and mud-volcano sites expelling methane-rich fluids from the seabed.  Thermal or solute-driven convection alone cannot explain such high velocities in low-permeability sediments.  Here, we demonstrate that in addition to  buoyancy, osmotic effects generated by the adsorption of methane onto the sediments can  create large overpressures, capable of recirculating seawater from the seafloor to depth in the sediment layer, then expelling it upwards at rates of up to a few hundreds of metres per year.  In the presence of global warming, such deep recirculation of seawater can accelerate the melting of methane hydrates at depth from timescales of millennia to just decades, and can drastically increase the rate of release of methane into the hydrosphere and perhaps the atmosphere.}

Methane is of great environmental importance as a greenhouse gas, and marine seeps are estimated to contribute some
37\% of the geological sources; likewise mud volcanism provides another 11--17\%  \cite{ar5}. 
Methane hydrates  are often found in the vicinity of submarine seeps and mud volcanoes \cite{seabed}, and the project of commercializing this energy source is under way. Concomitantly there is concern that anthropogenic climate change could destabilize hydrates, thereby potentially releasing large quantities of methane into the ocean and atmosphere \cite{buffett2000,archer2007,archer2009}. Thus there are both environmental and economic reasons to be interested in methane associated with seeps and mud volcanism. 

A mud volcano is a geological structure on land or in the oceans in which water plus fine particulates --- `mud' --- issues from a conduit typically topped by a conical hill with a crater. The emerging mud is generally accompanied by methane, both dissolved and, if the concentration exceeds the saturation concentration,  as bubbles \cite{kopf2002}.  A submarine seep, on the other hand, has similar fluid flow through the porous sediment constituting the seabed without the conical structure or open conduit. What physical forces drive such fluid flows? 

We find that both buoyancy and osmotic effects are present  in cold seeps and mud volcanism in which, rather than being a passive element, methane is its driving force.  Some researchers have suggested the importance of considering osmosis in seeps and mud volcanism \cite{soloviev, ginsburg}. Clays and shales are known to possess the semipermeability necessary for osmosis, associated with charge and pore-size effects \cite{neuzil,malusis}.  Moreover, methane hydrates frequently exist in the sediments around a cold seep or mud volcano \cite{kopf2002,seabed}.  As hydrate forms, the sediment plus hydrate becomes progressively less permeable \cite{seol2011,konno2015}.  However, other mechanisms involving adsorption and chemical reaction  can also produce significant osmotic pressures \cite{lee2014,derjaguin1972, derjaguin1965, manassero2003,biesheuvel,malusis2012,cc,brady}. Methane is shown to adsorb onto sediments \cite{sugimoto,ertefai,Ji2012,rexer2013}.  Hydrates form in pores under a wider range of conditions than in the bulk \cite{casco2015} and methane molecules adsorb onto the cages of methane hydrate during the hydrate growth process \cite{guo2009,zhong2014}. Thus, we propose below that given a supply of methane, a submarine cold seep or mud volcano can function as a geological instance of an osmotic pump \cite{neuzil,noy2004,barge2015}. 
We find cause for concern that this convective pump mechanism facilitates methane hydrate destabilization under anthropogenic climate change.

\section*{Results}
\subsection*{Liquid flow mechanisms in  seeps and mud volcanoes}

Consider the possible physical driving forces for a submarine seep in which salty water and methane are driven out of the seabed (Fig.~\ref{scheme}). The origin for this water can be either a reservoir beneath the seabed or seawater itself, recirculated within the porous seabed. 

One commonly discussed flow mechanism is the expulsion of pore water from sediments under compression, which can yield a speed of some $v_{\rm{sc}}=1.8\times 10^{-3}$~m~yr$^{-1}$ \cite{linke94}. This estimate provides a baseline with which to compare mechanisms. 
Buoyancy and osmotic forces are other possible driving mechanisms.
Let us consider 
quantitatively the processes for fluid transport in a cold seep in which the sea bed consists of a homogeneous porous medium, within which, at a given depth beneath the seafloor, we position (Fig.~\ref{scheme}a) an extended source of buoyancy, caused by either a thermal or a compositional difference, or (Fig.~\ref{scheme}b) a two-dimensional source of buoyancy, such as that at a continental margin.  Much previous work has explored the occurrence of flow focussing in overpressured heterogeneous sediments \cite{flemings1,flemings2}.
We further consider the case (Fig.~\ref{scheme}c) in which
 the sediment is heterogeneous, so that focussed flow may occur above a source of buoyancy or osmotic effects; for example, methane dissolved in warm water, the porous medium being semipermeable to this methane. We estimate the rate of flow of water induced in each case.  
 We should mention that there exists a further buoyancy-driven flow mechanism, that of liquid flow driven by buoyant bubbles \cite{O'Hara1995}, but this becomes dominant only with very large gas fluxes; that is, during the eruptive phase of mud volcanism, while here we concentrate on flows during the quiescent phase that subsists $\sim95$\% of the time \cite{kopf2002}.  
 
 The effectiveness of each mechanism of pumping fluid depends of course on the permeability of the sediment.
 Measurements  at seeps indicate values for inter-granular permeability in the range $10^{-18}$--$10^{-14}$ m$^2$ \cite{henry92}.  However, it has been suggested \cite{henry92, giambalvo2000, saffer2015} that bulk permeabilities are much higher, of the order of $10^{-12}$ m$^2$, owing to channelling of fluid through the sediment.  Below we assume a bulk permeability $k=10^{-12}$~m$^2$ and later extend our results to the range of  permeabilities $10^{-13}$--$10^{-11}$ m$^2$.

Positive buoyancy forces can arise from a temperature gradient that heats the water, or from solutal sources.  For example, heat is released when methane hydrates form and lower-density fresh water is generated when hydrates dissociate  (Fig.~\ref{scheme}a). 
When buoyancy from a geological process is released over a large area at depth in an otherwise quiescent, saturated porous medium, the less-dense fluid rises above the source (Fig.~\ref{scheme}a).  In a homogeneous sediment, the upward flow is essentially one-dimensional and we estimate (Methods) flow speeds $v_{\rm{est}} \sim 0.15$ m~yr$^{-1}$ and $v_{\rm{ess}} \sim 0.75$~m~yr$^{-1}$ for  thermal and solutal convection, respectively.

Consider now the release of heat over a long, thin area such as a continental margin. The rise of the warm fluid is now more localized, forming a line or two-dimensional plume (transverse view in Fig.~\ref{scheme}b). The plume increases in width as it rises, owing to viscous drag and to transverse heat conduction.  The rise velocity of the warm fluid decreases with height  as a consequence of cooling and the consequent reduction of buoyancy, while the volumetric flow rate increases on account of the increase in plume width. We consider a heat source of strength  $F_{\rm{t}}=25$~J~m$^{-1}$~s$^{-1}$ at  a depth $H=100$~m to be consistent with observations of surface heat fluxes of the order of 0.12~J~m$^{-2}$~s$^{-1}$ measured at the northern Cascadia accretionary sedimentary prism \cite{davis90}. The upward liquid speed at the seafloor is then estimated (Methods) as $v_{\rm{lst}} =0.25$~m~yr$^{-1}$, the half-width of the plume is $b_{\rm{lst} }=87$~m and the total flow rate per unit length is $Q_{\rm{lst}}=43$ m$^2$~yr$^{-1}$.  
 
A similar plume flow develops when less dense fluid is released in a localized region at depth, for example water liberated through methane hydrate dissociation \cite{henry96} or during the smectite--illite transformation \cite{pichon91}. In a given case, the
chlorinity of interstitial pore water sampled
at a seep site is about 0.95 times that of
seawater  \cite{henry92}, suggesting dilution of  low salinity water from depth with seawater by a factor of the order of 20. Also, heat balances over active seep areas taking into account measured background thermal  gradients have suggested dilution of the original water from depth with seawater by a factor of 6 to 30 \cite{pichon91}. Constraining the plume flow here to a dilution factor of 20 allows us to estimate (Methods) the original source fluid flow rate $Q_0=0.16$~m$^2$~yr$^{-1}$, so that the  the total flow rate per unit length at the seafloor  is $Q_{\rm{lss}}=3.1$ m$^2$~yr$^{-1}$.  We estimate the plume speed $v_{\rm{lss}} =0.13$~m~yr$^{-1}$ and half-width $b_{\rm{lss}} =12$~m.  This source flow rate is consistent with values quoted  for the release of water by the smectite--illite transformation and the hydrate layer \cite{pichon91}. 
 
		For comparison, consider now a buoyant flow in a heterogeneous porous medium under a seep (Fig.~\ref{scheme}c). We assume focussed flow directly under the seep area has increased the local permeability so that the main resistance to flow arises in the drawing of seawater from the surroundings into the rising seep plume. For an array of seeps spaced at approximately 50~m and radius $b_{\rm{s}}=2$~m \cite{Tryon1999,seabed}, we predict (Methods) a vertical velocity of $v_{\rm{ss}} \sim 59$~m~yr$^{-1}$, for a solutal source of buoyancy at depth $H=100$ m and a dilution ratio of 20.  The total flow rate in the convective cell is $Q_{\rm{ss}}=741$ m$^3$~yr$^{-1}$.  The effluent flux of methane averaged over the inflow area is approximately 0.0031 mol~m$^{-2}$~yr$^{-1}$.

Osmotic forces arise from a compositional gradient of fluid within a medium possessing a degree of semipermeability.  
Let us consider the flow induced by the osmotic pressure gradient associated with a release of fresh water saturated with methane at depth in the same  heterogeneous porous medium under a seep (Fig.~\ref{scheme}c). The water at the seafloor, above and beyond the seep region, is free from methane owing to continuous motion of ocean currents.  We expect seawater to flow downward from the seafloor into the porous sediment and towards the region with high concentration of methane. The methane in solution may diffuse into and through the surrounding seawater-saturated porous sediment.  A mixture of the source freshwater with methane and seawater will eventually rise in the form of an osmotic plume, exiting at the seafloor as a seep flow.
 We predict (Methods) a vertical velocity of $v_{\rm{so}} \sim 571$ m~yr$^{-1}$ for an osmotic source at depth $H=100$ m and a dilution ratio of 20.  The total flow rate in the convective cell is $Q_{\rm{so}}=5919$ m$^3$~yr$^{-1}$. The effluent flux of methane averaged over the inflow area is approximately 0.025 mol~m$^{-2}$~yr$^{-1}$.

These scaling and numerical results suggest that liquid efflux from the seabed in seep regions driven by osmotic pumping can be at least 10 times larger than in seeps with convection induced by salinity and thermal differences.  It is also far larger than reported velocities for water expulsion resulting from sediment compression.

Lastly, let us consider both buoyant and osmotic circulation in the vicinity of a fully-developed mud volcano with a fractal network of smaller channels at depth leading to the main central conduit, like a tree roots and trunk, above a source of solute or/and heat. Consider an idealized version of the geometry of the conduits: a vertical main channel, through which water with dissolved methane flows upwards from side feeder channels at depth and exits at the seafloor (Fig.~\ref{scheme}d).   Here, we consider the drawdown of water from the seabed caused by buoyant or osmotic pressure associated with a solute in solution in these conduits. Each conduit thus behaves like a buoyant or osmotic source considered above. Such network flows have been considered for many systems, from rivers in geology to the vascular system and the lungs in biology; our case corresponds to a directed spanning tree, the most efficient class of networks \cite{banavar1999}. We find that the speed of the flow in the main conduit is proportional to the total lateral area of all the feeder conduits (Methods). The combination of a buoyant or osmotic pump and a very small volume fraction of conduits within the sediment produces flow rates orders of magnitude larger than that in a homogeneous porous medium alone.

\subsection*{Comparison with field measurements }
 
The convective pump mechanism that we have demonstrated here functions as an amplifier of a small external source of buoyancy or dissolved methane into a large quantity of water that cycles through the seep or volcano. The question of the provenance of the water is a telling datum. The dominant contribution to the water issuing from submarine seeps and mud volcanoes is not  water from reservoirs under the seafloor, but seawater \cite{Paull1991,Tryon2001,mastalerz2007,nakayama2010}. 

In Fig.~\ref{methane_flux}  we present a comparison of in-situ measurements of liquid and methane fluxes from cold seeps \cite{linke94,torres2002} and mud volcanoes \cite{sauter2006,felden2010} with our theoretical predictions.   It is clear that thermal and solutal convection alone cannot explain the very high velocities measured at methane-rich sites with low permeability sediments. Contrariwise, osmosis induced by methane is likely to be the physical mechanism responsible for some of these high velocities, as revealed by the orange shaded ellipse.  Indeed, even with a conservative estimate of the osmotic effect,  osmotic pressure gradients are capable of producing very fast flows. The fluids vented at these seep sites had concentrations of dissolved methane in the range 0.6--126~$\mu$M. 
Osmosis is an efficient mechanism for producing fast localized flows owing to the relatively large pressure differences it generates.  While for the buoyant seep considered above the pressure driving the convective cell  is of the order of 4400~Pa, in the osmotic seep a pressure of 19400~Pa is achieved. 

Fluid flow into the seabed in the vicinity of a seep  has been observed; this flow pattern is difficult to understand from non-convective mechanisms. Measurements of downward speeds have been reported of 0.02--1.6~m~yr$^{-1}$ in the vicinity of methane vents in the Gulf of Mexico \cite{Solomon2008} and of  0.1--0.5 m~yr$^{-1}$ in Hydrate Ridge, Cascadia \cite{Tryon1999,Tryon2002}. These measurements may be compared with a predicted downward velocity at the seafloor surrounding the seep of approximately 3.1~m~yr$^{-1}$ for the osmotic flow, but only 0.38~m~yr$^{-1}$ for the buoyant case, so the faster downward flows, at least, should be owing to osmosis.

Our prediction of localized venting driven by osmosis is consistent with observations of seep regions in the  Gulf of Mexico involving cold and  dense saline effluent, where buoyancy can not drive the flow \cite{Coffin2008,Ruppel2005,Ruppel2008}.

There is at present just one measurement of flow rates in a mud volcano conduit, of 400 km~yr$^{-1}$ at a conduit of the H{\aa}kon Mosby mud volcano with radius 0.2~m (Mud volcano II \cite{sauter2006}; the other mud volcano measurement we plot, Mud volcano I \cite{felden2010}, corresponds not to a conduit, but to flow through a porous medium, as at a seep). This conduit flow measurement is compatible with our theoretical estimate for a mud volcano that predicts that the exit speed in a main conduit will be increased in proportion to the total lateral area of the network of conduits.  
Given a tiny volume fraction of 0.00001\% of the porous medium forming feeder conduits for a main conduit,  we predict such high-speed flows can be driven by both buoyancy or osmosis. This is a measure of how much more efficient than a seep a mud volcano is at pumping seawater.

\section*{Discussion}

Our model shows that recirculation of seawater within the seabed is rather greater and deeper than previously understood. While earlier models have considered convection of seawater in shallow layers of sediment of only a few metres depth \cite{pichon91,henry92,henry96}, here we have deduced that convection can extend to the depth of the source of buoyancy or dissolved methane.  Such flow can cool the sediment column  by up to $1.3 \times 10^7$ J~m$^{-2}$~yr$^{-1}$ per unit temperature difference between the seawater and the seep effluent.  Measurements of outflow temperatures at seeps indicate a temperature elevated  by 0 -- 5~K relative to  seawater {\cite{seabed}.  So we predict a maximum heat flux of $6\times 10^7$~J~m$^{-2}$~yr$^{-1}$ associated with convection in the sediment. We therefore envisage that recirculation of seawater to hundreds of metres depth will be problematic under conditions of climatic warming, as  buried methane hydrates  below seep and mud-volcano sites will be much more susceptible to destabilization than has been recognized up to now. 

Previous studies have estimated a timescale of millennia for conduction of heat from warmer seawater at the seafloor to affect the base of a hydrate layer at a few hundred metres depth and promote melting \cite{buffett2000,archer2007,archer2009}.  However, under enhanced heat transport by both buoyant and osmotic convection, we predict that the melting of hydrates could begin within timescales as short as 30 years. Such accelerated heat transport by convection will also increase the rate of melting of some hydrates by a factor of up to 100 compared to the heat conduction scenario previously studied (see Methods). 
The release of methane to the hydrosphere may thus occur much sooner and faster than previously thought. Such a continuous intense release of methane at the seabed will form a plume of rising methane bubbles that may reach the upper water column \cite{sauter2006,cardoso,domingos}.

It is challenging to assess what portion of the global inventory of methane hydrate, estimated as $1.8 \times 10^3$~Gt C \cite{Boswell2011}, might be susceptible to warming by the mechanism described here. For this, we need to combine oceanographic predictions for the warming of the upper few hundred metres of the ocean with hydrate stability studies. It is thought \cite{Ruppel2011}   that marine deepwaters on upper continental slopes (up to a few hundred metres depth), at the edge of the gas hydrate stability zone,  encompass some 3.5 \% of the global hydrate
inventory. It is also known \cite{archer2007} that shallow waters down to a few hundreds of metres respond to climate
change in roughly 10 years, while deep waters at 1--3 km take longer, 100--1000 years.
Based on these estimates, a maximum of about 3.5 \% of the global hydrate inventory ($\sim$ 60 Gt C) might be susceptible to warming by the mechanism proposed here within a timescale of a few decades.

\section*{Methods}

\subsection*{Notational note}
The subscripts $\rm{es}$, $\rm{lst}$, $\rm{lss}$, $\rm{so}$, and $\rm{ss}$ denote an extended source of buoyancy, a thermal margin plume, a solutal margin plume, a seep driven by osmosis and a seep driven by buoyancy, respectively. The subscripts `single' and `network' refer to a single conduit and a network of conduits in a mud volcano.

\subsection*{Uniform flow above an extended buoyancy source}
 
A buoyancy--viscosity balance suggests a velocity $v_{\rm{es}} \sim  k \Delta \rho  g / \mu$ driven by the density difference between the surrounding, less dense liquid $\Delta \rho$.  For a thermally-driven flow $\Delta \rho=\rho \beta_{\rm{t}} \Delta T = 2$~kg~m$^{-3}$ for a temperature difference $\Delta T = 10$~K \cite{davis90} and $\rho=1000$ kg~m$^{-3}$. 
For a solutal source of buoyancy, the maximum density difference driving the flow can be estimated to be smaller than $\Delta \rho=10$~kg~m$^{-3}$ taking into account salinity differences \cite{henry92} and heat absorbed during methane hydrate dissociation.
For the properties of water, we take the thermal expansion coefficient $\beta_{\rm{t}}=2 \times 10^{-4}$~K$^{-1}$ and the viscosity $\mu=1.8\times 10^{-3}$~kg~m$^{-1}$~s$^{-1}$. 

\subsection*{Thermal margin plume}

The vertical velocity at the centreline of a plume at a distance $H$ above the source is 
 $v_{\rm{lst}} =(k \beta_{\rm{t}} g F_{\rm{t}} /( \mu C_{\rm{p}} ))^{2/3} (3 / (32 \kappa_{\rm{m}} H))^{1/3}$  and the plume half-width is $b_{\rm{lst}} =(48 \mu C_{\rm{p}} \kappa_{\rm{m}}^2 H^2/( k \beta_{\rm{t}} g F_{\rm{t}}  ))^{1/3}$ \cite{turcotte}.
 The thermal diffusion coefficient of the saturated sediment is $\kappa_{\rm{m}}=10^{-7}$ m$^2$~s$^{-1}$; $g$ is the acceleration of gravity.   The specific heat capacity of water is $C_{\rm{p}}=4.2 \times 10^{3}$~J~kg$^{-1}$~K$^{-1}$. 

\subsection*{Solutal margin plume}

The plume velocity is
 $v_{\rm{lss}} =(k  \Delta \rho g Q_0/ \mu)^{2/3} (3 /( 32 D_{\rm{s}} H))^{1/3}$ and the plume half-width is $b_{\rm{lss}} =(48 \mu D_{\rm{s}}^2 H^2/( k  \Delta \rho g Q_0 ))^{1/3}$. We take the effective diffusivity of the solute causing the density difference, e.g., salt, in the porous medium as $D_{\rm{s}}=10^{-9}$ m$^2$~s$^{-1}$, and $\Delta \rho = 10$ kg~m$^{-3}$ as before.

 \subsection*{Pumping in a seep}
 
For pumping driven by osmosis, 
the porous medium behaves as only partially permeable to methane because a fraction of methane molecules of up to 0.6 are adsorbed \cite{linke94,ertefai} and later released by the sediment. This adsorption creates a change of momentum in the methane molecules that leads to a reflection coefficient of $\sigma_{0} <0.6$ \cite{ertefai,cc, biesheuvel, chapman}. For the pressure and temperature conditions in the seep data in Fig.~\ref{methane_flux}, the solubility of methane in water is in the range $c_0 \sim 0.10$--$0.23$~M \cite{duan}; we consider an intermediate value $c_0=0.156$~M.  A conservative estimate for the contribution of methane to osmosis is therefore $\sigma_0 c_0=0.008$~M, assuming that approximately 5\% of methane-molecule collisions with the sediment result in adsorption and later desorption \cite{sugimoto,Ji2012,rexer2013}. Seawater of course contains another solute, sodium chloride (other solutes found in  seep water  have much lower concentrations   \cite{mastalerz2007,nakayama2010}); we may neglect the osmotic effect of sodium and chloride ions because size-restriction effects in the sediment are very small for the permeabilities considered here\cite{neuzil,noy2004}. 
The semipermeability of the sediment to methane creates an osmotic pressure difference between the seawater and the methane-rich fluid released at depth of $p_0=\sigma_0 c_0 R T$; here, $R$ is the universal gas constant and $T \sim 283$~K is the temperature. A balance of osmotic and viscous forces gives the scale for the flow rate of seawater drawn into the osmotic plume  $Q_{\rm{so}} \sim   2 \pi a k p_0  /\mu$.  This flow will be channelled upwards towards the seafloor in a plume within the high permeability sediment column  below the seep. Consistent with seafloor observations, we assume a radius of $a=2$ m for this column  \cite{Tryon1999,seabed}.
The vertical velocity in the plume is $v_{\rm{so}}  \sim Q_{\rm{so}}/(\pi a^2)$. Numerical simulations neglecting the resistance in the upward flow compared to the downward and radial flow confirm this scaling, with a coefficient of 1.4 for seeps spaced at 50 m and a dilution ratio of the original source fluid of 20.  We estimate a vertical effluent speed $v_{\rm{so}} \sim 471$~m~yr$^{-1}$.  The vertical downward speed at the seafloor surrounding the seep is of the order of  $u_{\rm{so}}=3.1$ m~yr$^{-1}$.

For pumping driven by buoyancy, the pressure difference between the seawater and the hot or fresh fluid released at depth is $p_0=\Delta \rho g H$. A balance of buoyancy and viscous forces gives the scale for the flow rate of seawater drawn into the buoyant plume  $Q_{\rm{ss}} \sim   2 \pi a k p_0  /\mu$. Numerical simulations neglecting the resistance in the upward flow compared to the downward and radial flow confirm this scaling, with a coefficient of 0.35 for seeps spaced at 50 m and a dilution ratio of the original source fluid of 20.  We estimate a vertical effluent speed $v_{\rm{ss}} \sim 59$~m~yr$^{-1}$.  The vertical downward speed at the seafloor surrounding the seep is of the order of  $u_{\rm{ss}}=0.38$ m~yr$^{-1}$.

\subsection*{Pumping in a mud volcano}

A similar balance of buoyant/osmotic and viscous forces applies as for a seep, with the plume radius $a$ replaced by the conduit radius, $R_{\rm{c}}$.  Thus, for a single vertical conduit, the effluent speed $v_{\rm{single}} \sim Q_{\rm{so}} /(\pi R_{\rm{c}}^2)$ is to leading order independent of the conduit length.
For a network of conduits, with the simplifying assumptions that one conduit domain has no impact on the others and the viscous resistance to flow in the network is small compared to that in the porous medium, 
the flow rate is proportional to the  sum of the lateral surface areas $A_{\rm{i}}$ of all the individual conduits;  the exit velocity in the main conduit is then  $v_{\rm{network}}=v_{\rm{single}}\sum A_{\rm{i}}/A_{\rm{single}}$.  The exit speed thus depends on lengths and radii of the feeder conduits extending about the main conduit. 
For a single vertical conduit of radius $R_{\rm{c}}=0.2$~m, we predict a maximum speed of $v_{\rm{single}} \sim 3000$~m~yr$^{-1}$.  $v_{\rm{network}}$ can be several orders of magnitude larger than this: to achieve a hundredfold increase to $v_{\rm{network}}\sim 3 \times 10^5$~m~yr$^{-1}$, for example, the  lateral surface area of conduits needed is $\sum A_{\rm{i}} \sim 100\cdot 1000\cdot 0.2 \sim 10^4$~m$^2$, for $H = 1000$ m and $R_{\rm{c}} =0.2$ m. Assuming feeder-conduit radii $r_{\rm{c}} \sim 0.01$~m,  the corresponding volume of these conduits is  $\sim 10^2 $ m$^3$. The recirculation volume of the porous medium is $\sim 1000\cdot1000^2=10^9$ m$^3$. So, a hundredfold increase in the exit speed requires just a fraction of $10^{-7}$ of the recirculation volume of the porous medium to be conduits.

\subsection*{Propagation of a thermal signal from the seafloor}

The time of travel of a thermal signal from the seafloor to the base of the hydrate layer (where melting occurs) is approximately given by $H/u$, where $H$ is the distance from the seafloor to the base of the hydrate layer and $u$ is the superficial or Darcy vertical speed of the seawater moving downward in the sediment surrounding the seep or mud volcano.  $u$ was estimated from the scaling expressions above and confirmed by numerical simulation. 

The ratio of the heat flux associated with convection of seawater downward into the sediment and that associated with conduction is the P\'eclet number $Pe=uH/\kappa$.  We find $Pe\sim  100$ for the osmotic seep.

\subsection*{Data Availability Statement }
The authors declare that the data supporting the findings of this study are available within the article.

\bibliographystyle{naturemag}
\bibliography{hydrates}

\section*{Acknowledgements}
We thank Francesca Alvisi, Bruno Escribano, Diego Gonz\'alez, Ignacio Sainz-D\'{\i}az and Micol Todesco for useful discussions. SSSC acknowledges the financial support of the UK Leverhulme Trust project RPG-2015-002.  JHEC acknowledges the financial support of the Spanish MINCINN project FIS2013-48444-C2-2-P.

\section*{Author contributions}
Both authors contributed in an integrated fashion to this work.

\section*{Competing financial interests}
The authors declare no competing financial interests.

\pagebreak 

\begin{figure*}[t]
\begin{center}
\includegraphics*[width=\textwidth]{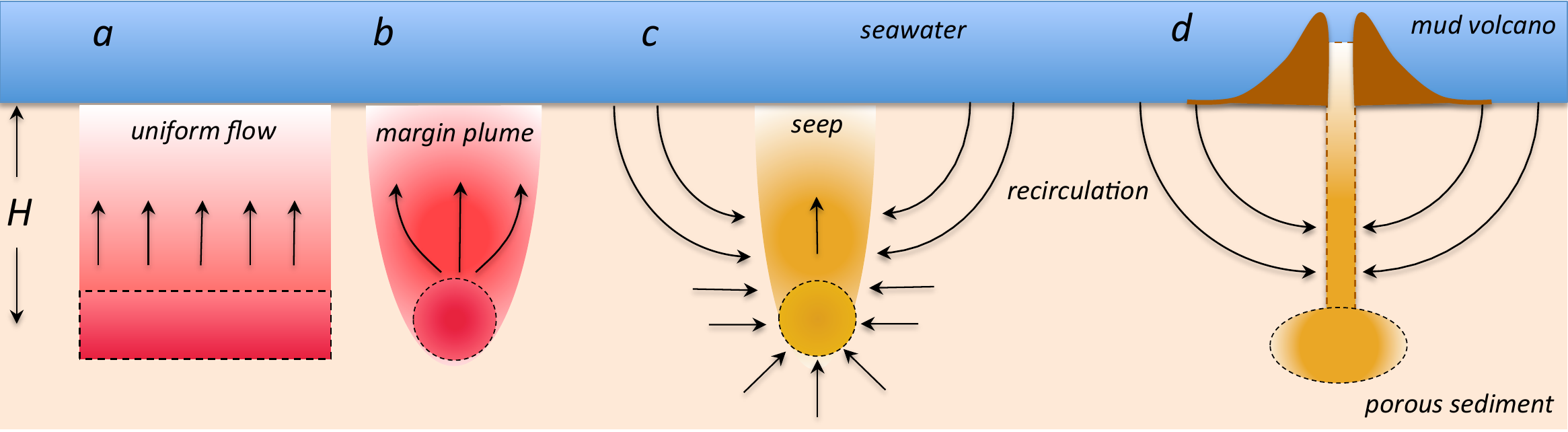}
\end{center}
\caption{{\bf Flow driven by buoyancy and osmotic sources in saturated porous sediment under the seafloor.}
(a) An extended buoyancy source in a homogeneous sediment, 
(b) a two-dimensional buoyancy source at a continental margin,
(c) a buoyant or osmotic pumping mechanism associated with a developed seep,
(d) buoyant or osmotic pumping in a fully-developed mud volcano.
}
\label{scheme}
\end{figure*}

\begin{figure*}[t]
\begin{center}
\includegraphics*[width=\columnwidth]{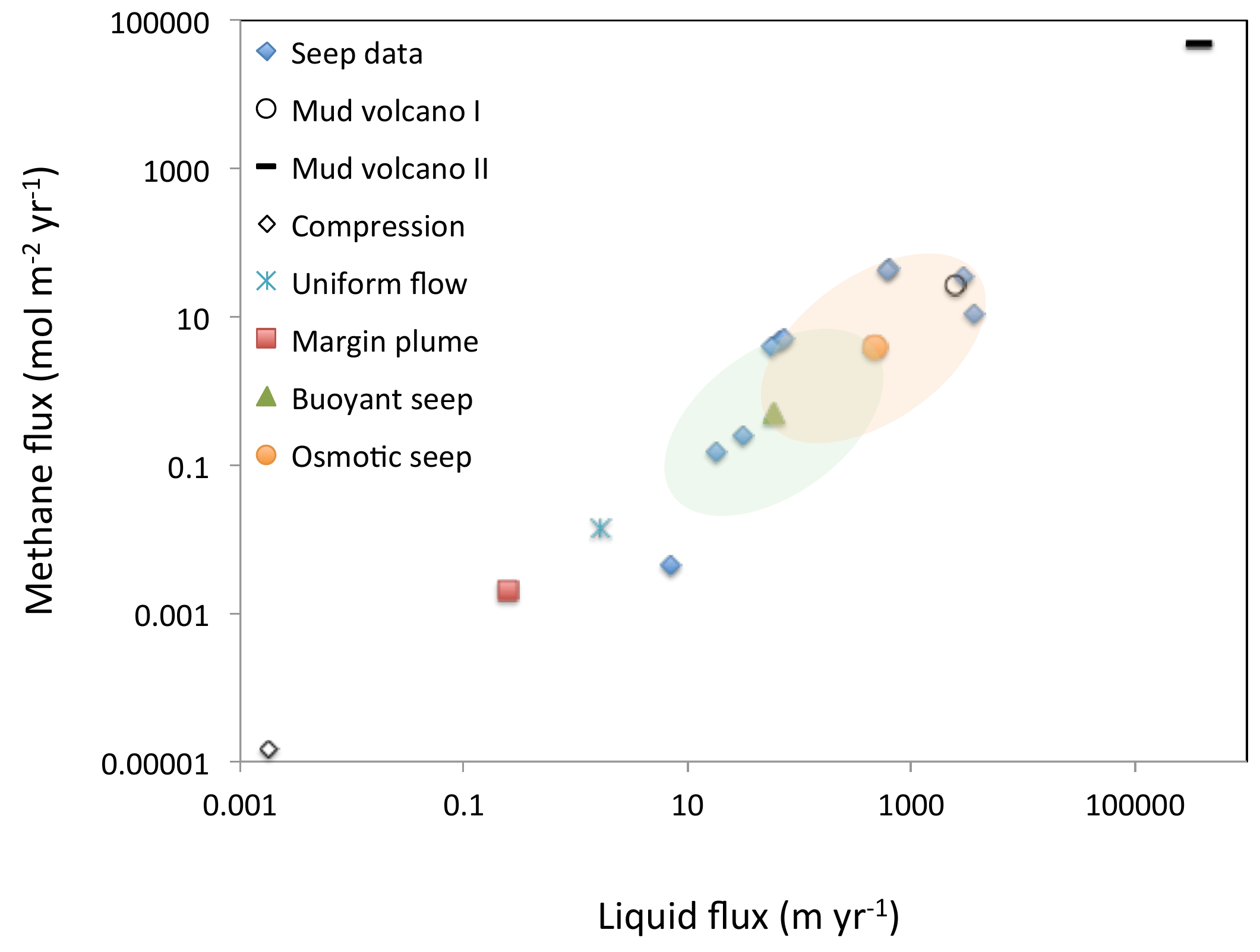}  
\end{center}
\caption{{\bf Dissolved methane flux plotted against liquid flux.} An osmotic mechanism can flow more methane and more liquid than competing mechanisms. Comparison of field measurements at seeps (\cite{linke94,torres2002}  and mud volcanoes (I \cite{felden2010},  II \cite{sauter2006}) with our theoretical predictions for a uniform source of solute and a margin heat plume, and a buoyant or osmotic plume in a developed seep. The predictions are for a sediment permeability of $10^{-12}$~m$^2$ and an exit methane concentration of 8~$\mu$M; the green and orange shaded ellipses represent the range of permeabilities $10^{-13}$--$10^{-11}$ m$^2$ (along the major axis) and methane concentrations  0.6--126~$\mu$M (along the minor axis) for a buoyant and osmotic seep flow, respectively. An estimate of efflux from sediment compression \cite{linke94} is shown as a baseline.
}
\label{methane_flux}
\end{figure*}

\end{document}